\documentstyle[preprint,aps]{revtex}
\newcommand{\be}{\begin{equation}}
\newcommand{\ee}{\end{equation}}
\newcommand{\bea}{\begin{eqnarray}}
\newcommand{\eea}{\end{eqnarray}}
\newcommand{\bm}{\begin{mathletters}}
\newcommand{\eml}{\end{mathletters}}
\newcommand{\pr}{\prime}
\newcommand{\oh}{{1\over 2}}
\newcommand{\tl}{\tilde}
\newcommand{\la}{\langle}
\newcommand{\ra}{\rangle}
\newcommand{\nn}{\\ \nonumber}

\begin{document}

\tighten

\begin{center}
\title
{\bf High spin proton emitters in odd-odd nuclei and shape changes \\}

\author
{D. S. Delion \\}
\address
{Institute of Physics and Nuclear Engineering, \\
Bucharest M\u agurele, POB MG-6, Romania \\}

\author
{R. J. Liotta, and R. Wyss \\}
\address
{KTH (Royal Institute of Technology) \\ AlbaNova University Center, \\
S-10691, Stockholm, Sweden \\}

\maketitle
%\date{\today}
\end{center}

\begin{abstract}

We present a formalism to describe proton emission from odd-odd nuclei
based on a scattering like approach.
Special emphasis is given to the case of transitions between states
with different deformations.
As an example we estimate the proton half-life of the odd-odd
nucleus  $^{58}$Cu.
Our calculations show that the change of deformation
in the decay process has a significant influence  on the half-life.
In addition, the angular momentum coupling of proton and
neutron orbitals  can result in an important $K$-hindrance
of the decay.
To fully account for the observed half-life in $^{58}$Cu,
one has to consider a shape mixing in the final state.
\end{abstract}

\vskip1cm


\vfill\eject

\rm

\renewcommand{\theequation}{\arabic{equation}}

The detection and study of nuclei on the proton drip line 
are being pursued intensely at present \cite{Woo97}. 
Many spherical as well as deformed nuclei 
in the region $50<Z<82$ have been shown to be proton emitters
\cite{Sel93,Pag94,Dav97,Bat98,Dav98,Bin99,Ryk99,Son99}.
Recently, an interesting example of prompt proton decay was reported:
the proton decay of the medium mass odd-odd nucleus $^{58}$Cu
which includes a transition from a highly deformed state into a
near spherical state, thus showing a pronounced shape change during
the transition from the mother to the daughter nucleus \cite{Rud98}.

The study of proton emission
are mainly based on a scattering like approach
\cite{Bug89,Kad96,Abe97,Dav00,Esb01} or on the R-matrix formalism
\cite{Mag99,Kru00,Bar00}.
A comparison of the two approaches was recently performed
in Refs. \cite{Dav00,Bia01}.
These studies were restricted to the case of decay from
odd-even to even-even nuclei. Only recently odd-odd nuclei have been
considered \cite{Fer01}. However,  in all cases it has been  assumed that
the shapes of the mother and the daughter nuclei are the same.

In this letter we present a formalism based on the scattering like
approach to study proton emission from deformed odd-odd nuclei including
transitions between excited states as well as 
different deformations between the mother and daughter nuclei. In addition,
we investigate the role played by the angular momentum orientation of the
odd particle in the intrinsic system.
As an example we will consider the case of
$^{58}$Cu, which life time has been established recently \cite{Rud01}.

We will present in some detail the case corresponding
to proton emission processes 
connecting two axially symmetric deformed nuclei.

The odd-odd mother wave function in the laboratory frame
is given by the ansatz \cite{Boh75}
\bea
\label{funi2}
\Psi^{J_i}_{M_iK_i}=
{1\over\sqrt{2(1+\delta_{K_i,0})}}
\left[
{\cal D}^{J_i}_{M_i,K_i}(\Omega)
\psi_{K_i}({\bf r}_{\pi}^{\pr},{\bf r}_{\nu}^{\pr})
+(-)^{J_i+K_i}
{\cal D}^{J_i}_{M_i,-K_i}(\Omega)\psi_{\tilde{K}_i}
({\bf r}_{\pi}^{\pr},{\bf r}_{\nu}^{\pr})
\right]~,
\eea
where $\Omega$ are the Euler angles and the normalised Wigner function
\bea
{\cal D}^{J_i}_{M_iK_i}(\Omega)\equiv\sqrt{2J_i+1\over 8\pi^2}
D^{J_i}_{M_iK_i}(\Omega)\eea
describes the rotation of the core. The prime coordinates
(e. g. ${\bf r}_{\pi}^{\pr}$) refers to the intrinsic frame.

The intrinsic wave function of the odd-odd (proton-neutron) system
is described by
\bea
\label{Nils}
\psi_{K_i}({\bf r}_{\pi}^{\pr},{\bf r}_{\nu}^{\pr})=
\Phi_{K_{\pi}}({\bf r}_{\pi}^{\pr})
\Phi_{K_{\nu}}({\bf r}_{\nu}^{\pr})~,
\eea
where the initial intrinsic spin projection is $K_i=K_{\pi}+K_{\nu}$.
The functions $\Phi$ are standard Nilsson wave functions
which can be written in terms of spherical components, i. e.
\bea
\label{func}
\Phi_{K_{\tau}}({\bf r}_{\tau}^{\pr})&=&
\sum_{lj}{g_{ljK_{\tau}}(r_{\tau})\over r_{\tau}}{\cal Y}_{ljK_{\tau}}
(\hat{r}_{\tau}^{\pr})~,
\nn
{\cal Y}_{ljK_{\tau}}(\hat{r}_{\tau}^{\pr})&\equiv&
\left[i^lY_l(\hat{r}_{\tau}^{\pr})\chi_{\oh}\right]_{jK_{\tau}}~,
~~~\tau=\pi,\nu~.
\eea
In order to evaluate the proton decay width we
express the mother wave function in terms of the 
coordinates of the outgoing proton in the laboratory frame.
By performing standard manipulations
the wave function (\ref{funi2}) becomes a superposition
of terms corresponding to the proton wave function coupled
to the rotational band built on the odd neutron state
with spin projection $K_{\nu}$, i. e.
\bea
\label{lab2}
\Psi^{J_i}_{M_iK_i}=
\sum_{J^\pr_f}
\sum_{lj} {g_{ljK_{\pi}}(r_{\pi}) \over r_{\pi}}
\phi_{J^\pr_flj}(\Omega,\hat{r}_{\pi},{\bf r}^{\pr}_{\nu})~,
%\varphi_{lj}(\Omega,\hat{r}_{\pi},{\bf r}^{\pr}_{\nu})~,
\eea
where 
\bea
\label{funl}
\phi_{J_flj}(\Omega,\hat{r}_{\pi},{\bf r}^{\pr}_{\nu})&=&
(-)^{j+K_i}\langle J_iK_ij-K_{\pi}|J_fK_{\nu}\rangle
\left[{\cal Y}_{lj}(\hat{r}_{\pi})
{\cal D}^{J_f}_{K_{\nu}}(\Omega)\right]_{J_iM_i}
\nn
&\times&
{1\over\sqrt{2(1+\delta_{K_i,0})}}
\left[\Phi_{K_{\nu}}({\bf r}_{\nu}^{\pr})+
(-)^{J_f+K_{\nu}}\Phi_{\tilde{K}_{\nu}}({\bf r}_{\nu}^{\pr})\right]~.
\eea
By using the orthogonality relation of these functions
\bea
\label{ort}
\langle
\phi_{J_flj}|\phi_{J_f^{\pr}l^{\pr}j^{\pr}}
\rangle=
\langle J_i,K_i;j,-K_{\pi}|J_f,K_{\nu}\rangle ^2
\delta_{J_jJ_f^{\pr}}
\delta_{ll^{\pr}}\delta_{jj^{\pr}}~,
\eea
the proton partial decay width to a fixed final state $J_f$
becomes \cite{Fro57}
\bea
\label{stre}
\Gamma_{J_f}=
\hbar v \sum_{lj}
\langle J_i,K_i;j,-K_{\pi}|J_f,K_{\nu}\rangle ^2
\lim_{r_{\pi}\rightarrow\infty}|g_{ljK_{\pi}}(r_{\pi})|^2~.
\eea
To obtain the radial parts of the wave functions, i. e. the functions
$g_{ljK_{\pi}}(r_{\pi})$ in Eq. (\ref{lab2}), one has to solve the 
corresponding Schr\"odinger equation, i. e.
\bea
\label{Ham}
\hat{H_{\pi}}\Psi^{J_i}_{M_iK_i}
=E_{\pi}\Psi^{J_i}_{M_iK_i}~,
\eea
where $H_{\pi}$ is the Hamiltonian describing the motion of the
emitted proton with positive energy
$E_{\pi}$.
We assume that the neutron is a spectator.
That is, the Hamiltonian $H_{\pi}$ will not affect the neutron
coordinates. 

One can solve Eq. (\ref{Ham})
by transforming to the intrinsic frame,
as in Eq. (\ref{funi2}). One then integrates over
the neutron coordinates and all angular coordinates
to obtain the system of equations corresponding
to the radial proton wave functions, i.e.
\bea
\label{sys}
{d^2g_{ljK_{\pi}}\over dr^2_{\pi}}&=&\lbrace{l(l+1)\over r^2_{\pi}}-
{2\mu\over\hbar^2}E_{\pi} \rbrace
g_{ljK_{\pi}}
\nn
&+&{2\mu\over\hbar^2}\sum_{l^{\pr}j^{\pr}}
\left[
\langle {\cal Y}_{ljK_{\pi}}|V_1|
{\cal Y}_{l^{\pr}j^{\pr}K_{\pi}}\rangle
g_{l^{\pr}j^{\pr}K_{\pi}}+
\langle {\cal Y}_{ljK_{\pi}}|V_2|
{\cal Y}_{l^{\pr}j^{\pr}K_{\pi}}\rangle
{dg_{l^{\pr}j^{\pr}K_{\pi}}\over dr_{\pi}}
\right]~,
\eea
Here $V_2$ is given by the non-spherical part
of the spin-orbit interaction \cite{Esb01}.

Our integration procedure using all possible channels
confirms that the initial state in the
odd-odd nucleus has a "stretched" configuration
$\pi g_{9/2}\otimes \nu g_{9/2}$ \cite{Rud98} with an amplitude 0.98.
For the initial state one therefore has $J_i=2j=9$.
As a result the system of equations (\ref{sys}) practically contains
only that component. That is, the outgoing proton escapes as a wave
containing all angular momenta but the contribution of the
configuration $(l,j)=(4,9/2)$ is clearly dominant.
Therefore, in our radial equations we match the external
solution with only that internal wave function.
We impose on the radial component $g_{ljK_{\pi}}(r_{\pi})$
outgoing boundary conditions at large distances, i. e.
\bea
\label{larg}
g_{ljK_{\pi}}(r_{\pi})=C_{ljK_{\pi}}{\cal H}^{(+)}_{ljK_{\pi}}(kr_{\pi})
\rightarrow  C_{ljK_{\pi}}\left[G_l(kr_{\pi})+iF_l(kr_{\pi})\right]~,
\eea
where $F_l(kr_{\pi})$ and $G_l(kr_{\pi})$ are the regular and irregular
Coulomb functions, respectively.
${\cal H}^{(+)}_{ljK_{\pi}}(kr_{\pi})$ denotes the numerical solution
found using backward integration.
The wave number is given by
\be
\label{dif}
k=\sqrt{2\mu E_{\pi}}/\hbar~.
\ee
The constant $C$ in Eq. (\ref{larg}) is obtained by matching
the external and internal solutions at some point $R$
\bea
\label{const}
C_{ljK_{\pi}}={\cal P}_{ljK_{\pi}}(R)/
|{\cal H}^{(+)}_{ljK_{\pi}}(kR)|~.
\eea
where ${\cal P}_{ljK_{\pi}}(R)$ is the proton formation amplitude.

Since
$\lim_{r_{\pi}\rightarrow\infty}|F_l(kr_{\pi})|^2+|G_l(kr_{\pi})|^2=1$
one can write the final expression for the proton decay width
(\ref{stre}) as
\bea
\label{wid}
\Gamma_{J_f}&=&\hbar v
\left[\langle J_i,K_i;j,-K_{\pi}|J_f,K_{\nu}\rangle
{\cal P}_{ljK_{\pi}}(R)/|{\cal H}^{(+)}_{ljK_{\pi}}(kR)|\right]^2~.
\eea

A similar formula was already presented in Ref.~\cite{Fer01}.

We will assume that the mother and daughter even-even cores can be
described within the BCS approach. As already mentioned the odd
neutron acts as a spectator.
The  proton-neutron interaction is included in
a phenomenological manner
by adjusting the proton and neutron pairing gaps
to their experimental values.
With these approximations
the formation factor ${\cal P}$ is computed as
the internal component multiplied by the amplitude
to find the final proton Nilsson state ($c^{\dag}_{ljK_{\pi}}$) in the
initial quasiparticle state ($a^{\dag}_{ljK_{\pi}}$), i.e.
\bea
\label{form}
{\cal P}_{ljK_{\pi}}(R)&=&g^{(int)}_{ljK_{\pi}}(R)
\langle BCS_f|c_{ljK_{\pi}}a^{\dag}_{ljK_{\pi}}|BCS_i\rangle~,
\eea
where
$g^{(int)}_{ljK_{\pi}}$ is the internal solution of the system
(\ref{sys}) and $|BCS_{i(f)}\rangle$ refers to the initial (final)
$^{57}$Ni core state.
One finally obtains
\bea
\label{form1}
{\cal P}_{ljK_{\pi}}(R)&=&g^{(int)}_{ljK_{\pi}}(R)u^{(f)}_{K_{\pi}}
{\cal U}_{K_{\pi}K_{\pi}}\langle BCS_f|BCS_i\rangle~.
\eea
Here $u^{(f)}_{K_{\pi}}(\tau)$ is the
proton BCS amplitude of the final nucleus.
The elements of the matrix ${\cal U}$ connect the quasiparticle
operators corresponding to the initial and final deformations
(for details see Appendix (E.3) of Ref. \cite{Rin80}).
The overlap integral of the two cores is given by
\bea
\label{over}
I\equiv\langle BCS_f|BCS_i\rangle &=&
|det[{\cal U}(\pi)]det[{\cal U}(\nu)]|^{1/2}~,
\nn
{\cal U}_{K_{\tau}'K_{\tau}}&=&
[u^{(f)}_{K_{\tau}'}u^{(i)}_{K_{\tau}}+
 v^{(f)}_{K_{\tau}'}v^{(i)}_{K_{\tau}}]
\langle \Phi^{(f)}_{K_{\tau}'}|
\Phi^{(i)}_{K_{\tau}}\rangle .
\eea
where $\Phi^{(i/f)}_{K_{\tau}}$ denote the single particle
Nilsson wave functions.
In Eq. (\ref{form}) we assume that the $K$-value for both the
odd-proton and neutron states are conserved during the decay process.

The proton wave function necessary to evaluate the
formation amplitude in Eq. (\ref{wid}), will be calculated by using
a Woods-Saxon mean field with universal parametrisation
\cite{Dud81}. We adjust the depth of the potential
in order to reproduce the $Q$-value.

The decay width (\ref{wid}), or the half-life $T=\hbar~ln2/\Gamma$,
should be independent of the radius $R$  in a region beyond
the nuclear surface.
This condition is automatically satisfied in our case
because the internal and external solutions satisfy
the same Schr\"odinger equation.

To analyse the $K$-dependence and to determine the shape
of the mother and daughter nuclei, we performed
total Routhian surface (TRS) calculations, where quadrupole
and hexadecapole deformations are minimized with respect to the
energy in the rotating frame of reference.
Pairing correlations are treated selfconsistently by
means of the Lipkin-Nogami method \cite{Sat94,Sat95,Wys95}.
The shape of the mother (initial) nucleus $^{58}$Cu is calculated to 
be prolate and axially symmetric.
The hexadecapole deformation is calculated to be $\beta_4=0.05$
while the quadrupole deformation parameter yields a value of
$\beta_2=0.35$.
The bandhead spin of the mother nucleus, assigned to be $J=9$,
corresponds to a predominantly $K=1/2$ configuration for both
protons and neutrons, that is fully rotationally aligned.

The initial $K$-value of $^{58}$Cu is given by
$K_i=|K_\pi \pm K_\nu|$ yielding a value of $K_i=1,~0$.
If one assumes that the daughter (final) nucleus $^{57}$Ni has 
the same prolate deformation as $^{58}$Cu the overlap is unity
and the half-live becomes $T \approx 0.8\times
10^{-16}~s$, which underestimates the
corresponding experimental value, i. e.
$T_{exp}\approx 0.5\times  10^{-12}~s$ \cite{Rud01},
by four orders of magnitude.

Actually the nucleus $^{57}$Ni is expected to be spherical. However,
our calculations yield the shape of the $J_f=9/2^+$
state to be oblate deformed,
with a deformation value of $\beta_2\approx -0.2$.
This oblate deformed state is approximately 600 keV lower in
energy as compared to the spherical minimum.
We thus expect the ground state of the
$J_f=9/2^+$ state to be oblate deformed and {\sl not} spherical.
Since the moment of inertia at oblate shapes is rather small, non
collective excitations can compete or mix into the corresponding
collective rotations. According to the TRS calculations,
the oblate collective minimum disappears with increasing angular momenta,
whereas the non-collective oblate minimum stays stable up to high frequencies.
The quadrupole deformation lifts the degeneracy of the $J_f=9/2^+$
multiplet, forming a distinct state with a $K$-value of $9/2$ as
ground state.
For collective rotational states one
expects a strongly coupled band involving  M1 transitions.
Indeed, the data on $^{57}$Ni  shows  a M1 structure
above the proposed $J_f=9/2^+$ state,
distinctly different from the rest of the level scheme \cite{Rud99}.
The first excited state with spin $11/2^+$ is also
$\approx$1~MeV lower in energy than
the corresponding excitations of the negative parity spherical states.
Although the structure resembles that of a strongly coupled band, it does not
follow a $J(J+1)$ pattern, pointing towards non collective excitations, in
agreement with the calculations.

Assuming that the mother nucleus is prolate and the daughter 
spherical the overlap (\ref{over}) becomes 
$I^2\approx 3.2 \times 10^{-2}$,
giving a half-life of $T\approx 0.24\times 10^{-14}~s$, which is still 
two orders of magnitude below the experimental value.
If one assumes the oblate shape discussed above for the daughter 
nucleus the half life increases further and
one obtains $I^2\approx 10^{-2}$ and 
$T\approx 0.7\times 10^{-14}~s$.
The change is not large as compared with the prolate-spherical 
transition assumed above.

In order to understand the discrepancy with the experimental halflife
$T$, we analysed in detail the different elements entering the
expression of the decay width. We thus noticed that
the $K$-values corresponding to the odd proton and neutron
which may couple to the initial projection $K_i$,
ranging from $K$=1/2 to $K$=9/2,
can change drastically the values
of the Clebsch-Gordan coefficient in Eq. (\ref{wid}).
In Fig. 1 we show the half-life computed according to the decay width
given by Eq. (\ref{stre}) as a function of the final (neutron)
spin projection in the intrinsic system $K_{f}=K_{\nu}$.
The solid lines correspond to decays where the mother and
daughter nuclei have the same deformation, i. e. where
$\beta_i=\beta_f=0.35$ and the overlap factor entering
Eq. (\ref{wid}) is virtually unity.
In addition to  $K_i=1,~0$ (upper and middle line) we also
considered the value $K_i=2$ (lower line). In the same figure
we show the effect of the change in deformation
due to the overlap, Eq. (\ref{over}), for prolate
and oblate shapes.
The dashed lines correspond to the case where
the daughter nucleus has an oblate deformation, i. e. where
$\beta_i=0.35$ and $\beta_f=-0.2$.

The oblate deformation has a profound influence on the
wave function of the daughter nucleus: Whereas for spherical
nuclei all $K$-values are degenerate and equally probable,
at oblate shapes the $K_{\pi}=9/2$ orbit of the $g_{9/2}$ subshell is
at the Fermi surface.
This implies, according to Eq. (\ref{wid}), that the transition
is forbidden since $K_\pi=1/2$ and  $K_\nu=9/2$ would
result in a vanishing Clebsch Gordan coefficient.
Thus, in addition to the {\sl shape hindrance} we
encounter a {\sl K-hindrance}.

Therefore the only possibility to account for the proton decay of $^{58}Cu$ is
to assume that the initial and final $K$-values are not pure.
This is in fact expected, since
the Coriolis interaction will induce $K$-mixing.
To further investigate the role of this interaction, we performed
additional particle plus rotor calculation for the initial and final nuclei
by using the Nilsson basis functions given by Eq. (\ref{func}), i.e.
\bea
\label{Corpsi}
\tl{\Phi}^{(i/f)}_{\nu}=
\sum_{K_{\nu}}d^{(i/f)}_{K_{\nu}}
\Phi^{(i/f)}_{K_{\nu}}~.
\eea
The odd neutron is not anymore a spectator
and its preformation factor is given by
\bea
\label{dd}
{\cal P}_{\nu}=
\la\tl{\Phi}^{(f)}_{\nu}\tl{\Phi}^{(i)}_{\nu}\ra=
\sum_{K_{\nu}}d^{(f)}_{K_{\nu}}d^{(i)}_{K_{\nu}}
\la\Phi^{(f)}_{K_{\nu}}|\Phi^{(i)}_{K_{\nu}}\ra~,
\eea
Our calculation shows that different eigenstates
with the same projection give a small contribution to 
${\cal P}_{\nu}$.

The effect of the Coriolis mixing is two-fold:

i) The initial and final $K_{\nu}$ will now acquire components of higher
and lower values, as can be seen in Table~1. From the same Table
it becomes clear that indeed the predominant component
in the initial prolate nucleus has $K_{\nu}=1/2$.

ii) The value of $K_i$ corresponding to
the band-head will change accordingly allowing additional coupling
possibilities.
The new term arising from the Coriolis mixed states implies
that the $K_{\pi}=\pm 1/2$ can couple
to $K_{\nu}=\mp 5/2$ yielding a
$K_i =2,~3$ state.
According to Table 1,
this is the only common component in both
initial and final nuclei. Higher $K$-values can be possible although
with even smaller amplitudes.

Then, assuming that the daughter nucleus is deformed with
$\beta_f=-0.2$ the calculated half-life for the case of
$K_i=2$  and $K_{f}=K_{\nu}=5/2$ (see Fig. 1)
becomes $T\approx 1.8\times 10^{-14}~s$.
The neutron preformation factor is
${\cal P}^2_{\nu}\approx 10^{-5}$. As a result
the lifetime acquires the value $T\approx 1.8\times 10^{-9}~s$,
which \it overestimates \rm the experimental value by three orders of
magnitude

Therefore, the experimental value would correspond to a decay in the
daughter nucleus where
the $9/2^+$ state has components in the wave function
corresponding to  oblate and spherical shapes, respectively.
In other words, shape fluctuations
due to the relative $\beta$-softness of the final nucleus
provides a suitable scenario to explain the actual lifetime.
Assuming that the  spherical and oblate
states in $^{57}$Ni are mixed, one obtains a lifetime
between the spherical
$T\approx 10^{-14}~s$ and oblate prediction $T\approx 10^{-9}~s$.
A simple estimate shows that the amplitudes of spherical and oblate
components should correspond to $X_{sph}\approx 0.1$ and
$X_{obl}\approx 0.9$, respectivelly, in order to reproduce
the experimental half-life.

In conclusion we have presented in this letter a formalism
to analyse proton decay in situations where the initial and final nuclei
are in excited states with different deformations.
We applied the formalism to study the decay of the odd-odd nucleus $^{58}$Cu 
and found that due to the difference of deformations between the
mother and daughter nuclei the calculated half-life increases
by two orders of magnitude, but still remains two orders of magnitude
below the experimental value.

Due to the prolate shape of the mother and
oblate shape of the daughter nucleus, the decay is K-hindered
and can proceed only via a very weak component
originating from the Coriolis mixing of initial and final states.
The half-life in this case becomes three orders of magnitude
larger than the experimental value.
A possible explanation of the measured half life is to assume
mixing of spherical and oblate shapes in the daughter nucleus, i. e. in
$^{57}$Ni.
The spherical component has a small amplitude $\approx 0.1$
but it is essential to describe the experimental half-life.

Given the large changes in life times, associated with the different
physical processes involved in the decay, implies of course a substantial
uncertainty in our analysis. Still, 
these unexpected features indicate that proton decay is a 
powerful tool to investigate
fine details of the intrinsic structure of nuclei at the drip line.

\vskip1cm

{\rm {\bf Acknowledgements} }

One of us (D S D ) thanks for the hospitality extended to him by the
Department of Physics of KTH, AlbaNova University Centre,
Stockholm, where part of this work was performed. This work
is supported by the Swedish Science Council.

\vskip1cm

\centerline{\bf Figure captions}

\vskip0.5cm

{\it Fig. 1.}

Half-life of $^{58}$Cu as a function of
the final spin projection in the intrinsic system $K_f=K_{\nu}$.
The solid lines correspond to transitions without deformation change,
i. e. with  $\beta_f=\beta_i=0.35$.
The dashed lines correspond to $\beta_i=0.35$ and
an oblate final shape with $\beta_f=-0.2$. The labels of the lines to
the right of the figure are the corresponding values of $K_i$.

%0.986, 0.212, 0.053
%0.901, 0.405, 0.089
%0.869, 0.172, 0.067, 0.01, 0.004

\vskip1cm

%\begin{table}[p]
\begin{center}
{\bf Table. 1}
\vskip0.5cm
{\it Amplitudes defined by Eq. (\ref{Corpsi}) corresponding to
the different $K_{\nu}$-values for the $9/2^+$
state at prolate and oblate shapes, respectively.}
\vskip0.5cm
\begin{tabular}{cccccc}
\hline
$K_{\nu}$ & 1/2 & 3/2 & 5/2 & 7/2 & 9/2\\
\hline
$d^{(i)}_{K_{\nu}}$ &0.901 &0.405&0.089&-&-\\
$d^{(f)}_{K_{\nu}}$ & - & - &0.053& 0.212& 0.986 \\
%oblate$\gamma=-50^\circ$&0.004&0.01&0.067&0.172&0.869\\
\hline
\end{tabular}
\end{center}
%\end{table}

%\begin{thebibliography}{99}

%\end{thebibliography}

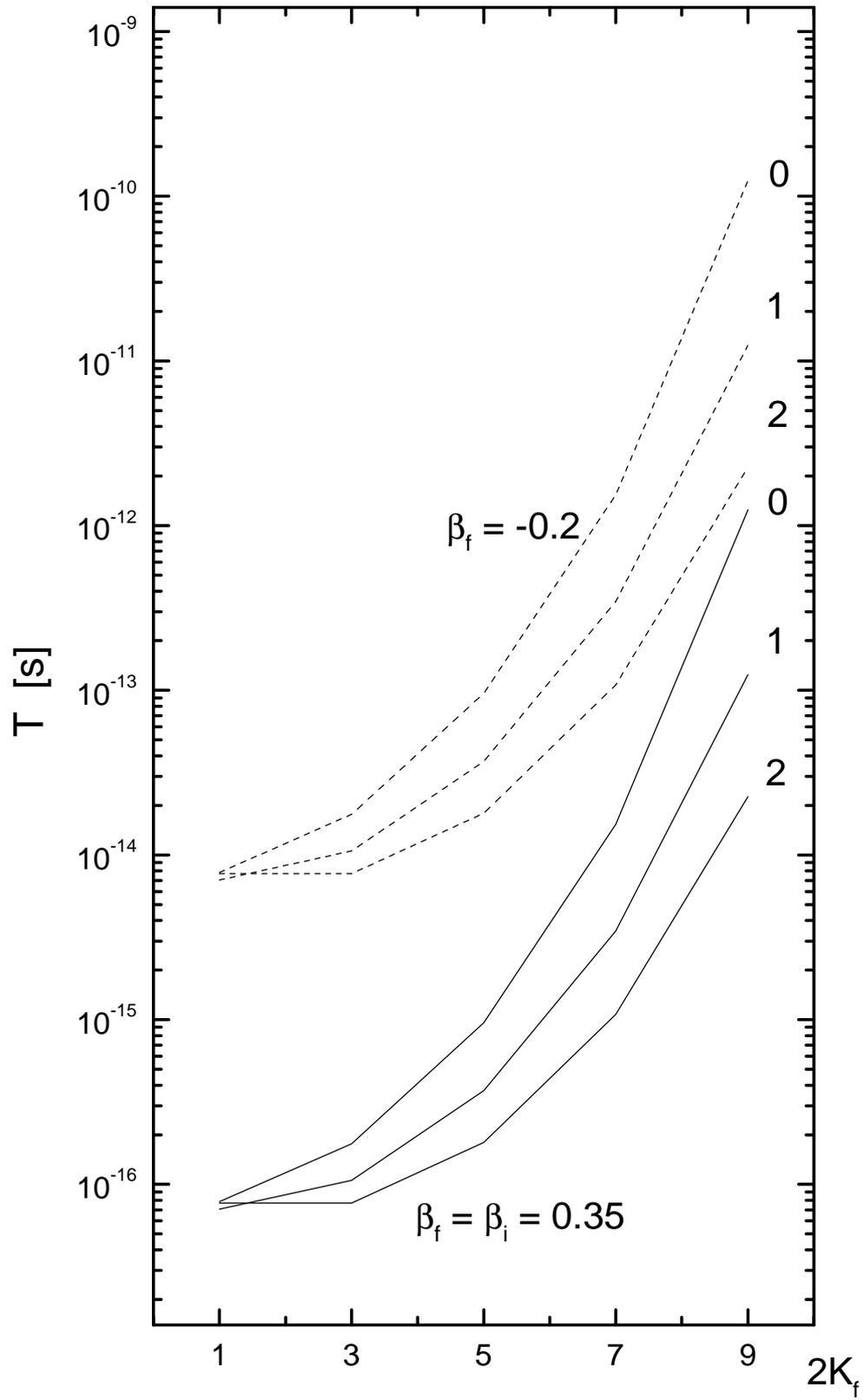
\end{document}